\documentclass[preprint,showpacs,preprintnumbers]{revtex4}
\usepackage{amssymb}
\usepackage{amsmath}
\usepackage{graphicx}
\usepackage{dcolumn}
\usepackage{bm}

\input{tcilatex}
\begin{document}

\title{Fidelity and Fidelity Susceptibility of Pulses in Dispersive Media}
\author{Li-Gang Wang$^{1,2}$ and Shi-Jian Gu$^{1}$}
\affiliation{$^{1}$Department of Physics, The Chinese University of Hong Kong, Shatin, N.
T., Hong Kong\\
$^{2}$Department of Physics, Zhejiang University, Hangzhou, 310027, China}

\begin{abstract}
Motivated by the growing importance of the fidelity and fidelity
susceptibility (FS) in quantum critical phenomena, we use these concepts to
describe the pulse propagation inside the dispersive media. It is found that
there is a dramatic change in the fidelity and the FS of the pulse at a
critical propagation distance inside a dispersive medium, and whether such a
dramatic change for a light pulse occurs or not strongly depends on both the
dispersive strength of the media and the pulse property. We study in detail
about the changes of the fidelity and the FS for both a smooth and a
truncated Gaussian pulse through the abnormal and normal dispersive media,
where the group velocities are well defined. Our results show that both the
fidelity and the FS could be very useful to determine whether the pulse is
completely distorted or not at the critical distance, therefore it would be
very helpful to find the maximal effective propagation region of the pulse's
group velocity, in terms of the changes of the pulse's fidelity and FS.
\end{abstract}

\maketitle

\section{INTRODUCTION}

Recently, to control the pulse propagation inside various media has
attracted intensive attention due to potential applications in information
processing and communication systems. It is now well known that the group
velocity of a light pulse can be controlled by tailoring the dispersive
properties of various materials \cite{ReviewChiao,Zubairy2005}. Many recent
experiments \cite{Chu1982,WANG2000,Pesala2006,Jiang2007,Song2008} have
demonstrated that the group velocity can be larger than the speed of light
in vacuum, and even become negative in various situations, such as anomalous
dispersive media and photonic band-gap materials. It has been discovered
also that the group velocity could become extremely slow in normal
dispersive media, such as an ultracold atomic gas \cite{Hau1999} and
solid-state materials \cite{Kasapi1995,Xiao1995}. However, due to the
effects of materials' dispersion and absorption (or gain), the light pulse's
shape always suffers an inevitable distortion, which is directly related to
the propagation distance of the pulse inside a given medium. Then one may
wonder: is there any limitation on the propagation distance for a light
pulse inside the dispersive media? And how to quantitatively determine the
maximal effective propagation region for a given light pulse without
distorting its original shape? Since it is very difficult to quantitatively
measure the degree of the pulse distortion, to the best of our knowledge,
the above questions have not been answered.

The concept of fidelity is originally defined in the field of quantum
information theory. It measures the similarity between two quantum states 
\cite{Nilesen2000book,ReviewSteane}. Recently, this concept has been used to
characterize quantum phase transitions in various quantum many-body systems 
\cite{Quan2006,Zanardi2006,Zanardi2007,Buonsante2007,Zhou2008,GuDFFS}. It
has been found that a dramatic change of a quantum state should result in a
great change in fidelity across a critical point. In prior studies, the
quantum fidelity is used to depict the distance between two slightly
different states corresponding to two slightly different values of the
parameters \cite{Zanardi2006} in the parameter space. Thus fidelity reflects
a response of a quantum state to a change of the driving parameter. In order
to study the fidelity, one should in principle choose a small distance in
the parameter space. This distance is arbitrary and not so important in
relevant studies. It was later pointed out that leading term of fidelity,
called fidelity susceptibility (FS), plays a central role \cite%
{Gu2007a,Zanardi2007PRL,Gu2008a,Gu2008b} in the fidelity approaches to
quantum phase transitions.

In this paper, we use the fidelity and the FS of a pulse wavepacket and try
to answer the above questions in terms of the changes of fidelity and FS. We
find that the fidelity and the FS are very useful to determine the range of
the effective propagation distance for a light pulse almost preserving its
original shape before a strong distortion of the pulse wavepacket happens
inside dispersive media.

This paper is organized as follows: In Sec. II, we simply present a brief
proof on the equivalent description of a light pulse wavepacket in both the
temporal and frequency domains. In Sec. III, we define the concepts of the
fidelity and the FS for a light pulse propagating through a dispersive
medium in the frequency domain. In Sec. IV, we study numerically the changes
of the fidelity and the FS of both a smooth and a truncated Gaussian pulses
inside the different types of dispersive media. We also analyze the
influence of both the dispersive media and the pulse itself on the changes
of the fidelity and the FS. Finally, our conclusions are given in Sec. V.

\section{An equivalent description of a pulse wavepacket between the
temporal and frequency domains}

It is well-known that, within a linear response theory and a moving time
frame $\tau $ ($\tau =t-z/c$, here $c$ is the light speed in vacuum), the
evolution equation of the wavepacket $\Psi \left( z,\tau \right) $ for a
light pulse propagating inside a dispersive medium can be expressed as \cite%
{LGWANG2003,Zhu2003} 
\begin{equation}
\Psi \left( z,\tau \right) =\frac{1}{\sqrt{2\pi }}\int \Psi \left(
0,t_{1}\right) G(\tau -t_{1},z)dt_{1},  \label{INOUT1}
\end{equation}%
where $\Psi \left( 0,t_{1}\right) $ is the initial wave packet of a light
pulse at position $z=0$ in the temporal domain, and $G(\tau -t_{1},z)$ is
the medium's Green function given by%
\begin{equation}
G(\tau -t_{1},z)=\frac{1}{\sqrt{2\pi }}\int \exp [-i\omega (\tau
-t_{1})]\exp \left\{ \frac{i\omega \lbrack n(\omega )-1+i\kappa (\omega )]z}{%
c}\right\} d\omega .  \label{GreenFun2}
\end{equation}%
Here the real functions $n(\omega )$ and $\kappa (\omega )$ are related to
the complex refractive index, satisfying the relation:%
\begin{equation}
n(\omega )+i\kappa (\omega )=\sqrt{\epsilon (\omega )},
\end{equation}
where $\epsilon (\omega )$ is the dielectric function of a non-magnetic
medium. From Eqs. (\ref{INOUT1}) and (\ref{GreenFun2}), we can find the
relation:%
\begin{equation}
\psi \left( z,\omega \right) =\psi \left( 0,\omega \right) \exp \left\{ 
\frac{i\omega \lbrack n(\omega )-1+i\kappa (\omega )]z}{c}\right\}
\label{Spectrum1}
\end{equation}%
in the frequency domain, where $\psi \left( z,\omega \right) $ is the
spectral wavepacket in the frequency domain, $\Psi \left( z,\tau \right) $
an inverse Fourier transformation of the wave packet in the temporal domain.
In Eq. (\ref{Spectrum1}), the function $\psi \left( 0,\omega \right) $ is
given by%
\begin{equation}
\psi \left( 0,\omega \right) =\frac{1}{\sqrt{2\pi }}\int \Psi \left(
0,t_{1}\right) \exp (i\omega t_{1})dt_{1},  \label{inputspect}
\end{equation}%
which is actually a transformation for the initial wave packet $\Psi \left(
0,t_{1}\right) $ from the temporal domain into the frequency domain. From
Eq. (\ref{Spectrum1}), we can find that in the moving time frame the
spectral wave packet $\psi \left( z,\omega \right) =\psi \left( 0,\omega
\right) $ keeps unchanged in the vacuum [$n(\omega )=1$ and $\kappa (\omega
)=0$]. This result is in a good agreement with our common sense that the
light temporal wavepacket in vacuum is unchanged in the moving time frame.
Therefore, for the description of a light wave packet, the expression $\Psi
\left( z,\tau \right) $ in the temporal domain is totally equivalent to the
expression $\psi \left( z,\omega \right) $ in the frequency domain.

\section{Fidelity and Fidelity Susceptibility in Dispersive Media}

Now we would like to use the wavepacket in the frequency domain to define
the fidelity and the FS of a light pulse when it propagates inside a
dispersive medium. Generally, in a dispersive medium [$n_{r}(\omega )\neq 1$
and $\kappa (\omega )\neq 0$], the pulse spectral wavepacket changes due to
the medium's dispersion and absorption (or gain) as it propagates over a
distance $z$. Therefore the fidelity of a light pulse wave packet must
change as it propagates through a medium.

We consider the propagation of a light pulse passing through a non-magnetic
dispersive medium with its dielectric function%
\begin{equation}
\epsilon (\omega )=1+\chi _{M}(\omega ),
\end{equation}%
where $\chi _{M}(\omega )$ is the linear susceptibility of the dispersive
medium, such as the two-level \cite{Kim2003}, three-level \cite%
{WANG2000,Agarwal2001}, or multilevel atomic gas media \cite%
{Sahrai2004,Mahmoudi2006,LGWANG2008}. In the frequency domain, the wave
packet of a light pulse at position $z$ is $\psi \left( z,\omega \right) $.
After it passes through a medium with a length $\delta z$, the pulse
spectral wave packet becomes $\psi \left( z+\delta z,\omega \right) $.
Following the definition in Ref. \cite{Zanardi2006}, the fidelity is defined
as the modul of the overlap of two quantum states. In our cases of classical
wavepacket, the fidelity can be also defined as the overlap between two
slightly different wavepackets $\psi \left( z,\omega \right) $ and $\psi
\left( z+\delta z,\omega \right) $ in the frequency domain, that is,%
\begin{equation}
F(z,\delta z)=\frac{1}{N(z,\delta z)}\left\vert \int \psi \left( z,\omega
\right) \cdot \psi ^{\ast }\left( z+\delta z,\omega \right) d\omega
\right\vert ,  \label{Fidelity1}
\end{equation}%
where $N(z,\delta z)$ is a real normalized function, which is given by%
\begin{equation}
N(z,\delta z)=\left[ \int \left\vert \psi \left( z,\omega \right)
\right\vert ^{2}d\omega \int \left\vert \psi \left( z+\delta z,\omega
\right) \right\vert ^{2}d\omega \right] ^{1/2}.  \label{NormalizedConstant1}
\end{equation}%
Now the propagation distance $z$ becomes a driving parameter. Combining with
Eq. (\ref{Spectrum1}), we can see from Eq. (\ref{Fidelity1}) that the
fidelity $F$ is exactly equal to one in the vacuum [$n_{r}(\omega )=1$ and $%
\kappa (\omega )=0$]. This indicates that \textit{the light pulse does not
change in vacuum}. At the same time, from Eq. (\ref{Fidelity1}), we can see
that as $\delta z$ approaches zero, $F$ still goes to one even if the pulse
propagates inside the dispersive medium.

In general, when a pulse propagates inside the dispersive medium, $F$ is a
well-defined monotonical and decreasing function with respect to $\delta z$.
Therefore, we can expand Eq. (\ref{Fidelity1}) by Taylor's expansion, (for
example, see Refs. \cite{Gu2007a,Gu2008b})%
\begin{eqnarray}
\left. F(z,\delta z)\right\vert _{\delta z\rightarrow 0} &=&F(z,\delta
z=0)+\left. \frac{\partial F(z,\delta z)}{\partial \delta z}\right\vert
_{\delta z=0}\delta z  \label{Define1} \\
&&+\frac{1}{2}\left. \frac{\partial ^{2}F(z,\delta z)}{\partial ^{2}\delta z}%
\right\vert _{\delta z=0}(\delta z)^{2}+O((\delta z)^{3}), \\
&=&1-\left[ -\frac{1}{2}\left. \frac{\partial ^{2}F(z,\delta z)}{\partial
^{2}\delta z}\right\vert _{\delta z=0}\right] (\delta z)^{2}+O((\delta
z)^{3}).  \label{Define2}
\end{eqnarray}%
We can analytically prove the first derivative in Eq. (\ref{Define1}) to be
zero given the parameter $\delta z\sim 0$, therefore the most relevant term
in determining the fidelity of a light pulse is its second derivative.
According to the response theory, the coefficient term before $(\delta
z)^{2} $ actually defines a response of fidelity to a small change in the
propagation distance. Following with Ref. \cite{Gu2007a}, we can introduce
the FS for a light pulse as follows,%
\begin{equation}
\chi _{F}(z)=-\left. \frac{2\ln F}{\left( \delta z\right) ^{2}}\right\vert
_{\delta z=0}.  \label{FS1}
\end{equation}%
With Eq. (\ref{Fidelity1}), the FS in a dispersive medium can be expressed as%
\begin{equation}
\chi _{F}(z)=-\frac{M^{2}(z)}{P^{2}(z)}+\frac{2M(z)\func{Re}[Q(z)]}{P^{2}(z)}%
+\frac{U(z)}{2P(z)}-\frac{\func{Re}[X(z)]}{P(z)}-\frac{\left\vert
Q(z)\right\vert ^{2}}{P^{2}(z)},  \label{FS2}
\end{equation}%
where the functions $P(z)$, $M(z)$, $Q(z)$, $U(z)$, and $X(z)$ are defined
as follows%
\begin{equation}
P(z)\equiv N(z,\delta z=0),
\end{equation}%
\begin{equation}
M(z)\equiv -2\int \frac{\omega }{c}\kappa (\omega )\left\vert \psi \left(
z,\omega \right) \right\vert ^{2}d\omega ,
\end{equation}%
\begin{equation}
Q(z)\equiv \int \left\{ \frac{i\omega }{c}[n(\omega )-1]-\frac{\omega }{c}%
\kappa (\omega )\right\} \left\vert \psi \left( z,\omega \right) \right\vert
^{2}d\omega ,
\end{equation}

\begin{equation}
U(z)=4\int \frac{\omega ^{2}}{c^{2}}\kappa ^{2}(\omega )\left\vert \psi
\left( z,\omega \right) \right\vert ^{2}d\omega ,
\end{equation}

\begin{equation}
X(z)\equiv \int \left\{ \frac{i\omega }{c}[n(\omega )-1]-\frac{\omega }{c}%
\kappa (\omega )\right\} ^{2}\left\vert \psi \left( z,\omega \right)
\right\vert ^{2}d\omega .
\end{equation}%
From Eq. (\ref{FS2}), we can readily obtain the change of the FS in the
dispersive medium. For the vacuum, since $n(\omega )-1=0$ and $\kappa
(\omega )=0$ for all $\omega $, one can easily verify that the value of $%
\chi _{F}(z)$ is always equal to zero, which means that the status of the
light pulse does not change in the vacuum. This conclusion is in agreement
with that from Eq. (\ref{Fidelity1}). In the following discussion, we will
show that the change of $\chi _{F}(z)$ indicates the different response
properties of the different dispersive media for a light pulse.

\section{Numerical Results and Discussions}

In this section, we would like to discuss the changes of the fidelity and
the FS of a light pulse in two types of non-magnetic dispersive media, such
as double-gain anomalous dispersive media and double-absorptive normal
dispersive media, where the group velocity has been well defined and could
be superluminal or subluminal. The linear susceptibility of the media in our
cases is a form of double Lorentz oscillators \cite{WANG2000},%
\begin{equation}
\chi _{M}(\omega )=\frac{M}{\omega -\omega _{0}-\Delta +i\gamma }+\frac{M}{%
\omega -\omega _{0}+\Delta +i\gamma },  \label{Medium1}
\end{equation}%
where $M\ $is proportional to the strength of the two oscillators, $\Delta $
the frequency detuning, $\gamma $ the damping rate of the two oscillators,
and $\omega _{0}$ the pulse carried frequency. Such a linear susceptibility $%
\chi _{M}(\omega )$ may describe a three-level atomic system with two
closely placed Raman gain peaks \cite{WANG2000}. Near $\omega _{0}$, when $%
M<0$, the medium is an absorptive normal dispersive medium, and when $M>0$,
it becomes a gain abnormal dispersive medium. In the following two
subsections, we will show that the changes of the fidelity and the FS are
strongly affected by different types of dispersive media and the property of
the pulse itself. From the fidelity and the FS of the pulse inside the
dispersive medium, one can determine the effective propagation distance of a
pulse.

\subsection{A smooth Gaussian pulse}

First, let us consider the propagation of a \textit{smooth} Gaussian pulse
passing through a dispersive media. For a smooth Gaussian pulse, its
wavepacket in the time domain is assumed to be in the Gaussian form%
\begin{equation}
\Psi \left( 0,t_{1}\right) =A_{0}\exp \left[ -\frac{t_{1}^{2}}{2T_{0}^{2}}%
\right] \exp [-i\omega _{0}t_{1}]
\end{equation}%
at the incident end $z=0$, where $A_{0}$ is a constant, $T_{0}$ is the
temporal half-width of the Gaussian pulse, and $\omega _{0}$ is the carried
frequency. Its corresponding pulse spectral wavepacket in the frequency
domain could be readily obtained as follows \cite{LGWANG2003}%
\begin{equation}
\psi (0,\omega )=\frac{T_{0}A_{0}}{2\sqrt{\pi }}\exp \left[ -\frac{%
T_{0}^{2}(\omega -\omega _{0})^{2}}{2}\right] .
\end{equation}%
Using Eq. (\ref{Spectrum1}), we can find the pulse spectral wavepacket at
any position inside dispersive media. Then using Eq. (\ref{Fidelity1}), we
can obtain the fidelity $F(0,z)$ between the initial pulse and the pulse at
a certain distance. At the same time, using Eq. (\ref{FS2}), we can find the
change of the FS for optical pulses inside dispersive media. Without loss of
generality, we take $\omega _{0}/2\pi =3.5\times 10^{14}$Hz, and pulse
temporal width $T_{0}=1.2\mu $s in the next numerical calculations.

Figure 1 shows the changes of the fidelity $F(0,z)$ and the FS $\chi _{F}(z)$
for a smooth Gaussian pulse propagating through a double-gain anomalous
dispersive medium with $M>0$ in Eq. (\ref{Medium1}). From Figs. 1(a) and
1(c), the fidelity is a monotonical, decreasing function. Initially the
fidelity smoothly decreases with respect to the propagation distance $z$,
and after a critical propagation distance $z_{c}$ we find that the fidelity
suddenly decreases very quickly. Correspondingly, close to the sudden
decrease of the fidelity, there is a peak in the curve of the FS as shown in
Figs. 1(b) and 1(d). In Figs. 1(a) and 1(b), we can see that, as the
parameter $M$ decreases, that is, as the anomalous dispersion strength of
the medium decreases, the sudden decrease of the fidelity has a larger
critical $z_{c}$ and correspondingly the peak on the curve of the FS also
appears at a longer propagation distance. There is a similar effect when we
change the frequency detuning $\Delta $. For a larger $\Delta $, which means
a weaker anomalous dispersion near $\omega _{0}$, both the sudden decrease
of the fidelity and the peak of the FS will appear at a longer propagation
distance.

Figure 2 show the normalized temporal intensities in the moving time frame $%
\tau $ at different positions, which are denoted on the curve A of Fig.
1(d), as a smooth Gaussian pulse propagates through an anomalous dispersive
medium. We can that the pulse almost keeps the same shape from Figs. 2(1) to
2(3) except for a small distortion. During this process, the FS is very
small. As the propagation distance reaches the peak's position of the FS,
the pulse's shape suffers a great change, and the intensity of the pulse's
distortion has almost the same order with that of the pulse itself [see Fig.
2(4) to 2(5)]. After the FS's peak, the output pulse is totally distorted
and has a completely different property from its initial pulse.

From a physical point of view, we know that the fidelity is a measurement of
the overlap between the initial and output pulse spectral wavepackets. From
the comparison between Figs. 1(a, c) and Figs. 1(b, d), we conclude that,
when the fidelity decreases smoothly within a short propagation distance,
the output pulse spectral wavepacket only suffers a \textit{quantitative}
deviation from its initial pulse spectral wavepacket. That is the output
pulse is almost the same as the input one so that the FS changes very slow
[almost a small constant before the peak of the FS, see in Figs. 1(b) and
1(d)]. However, when the fidelity has a sudden change at a critical distance 
$z_{c}$, actually there is \textit{a dramatic change} in the output pulse
spectral wavepacket. This tells us that the output pulse has a \textit{%
qualitative} change from its initial pulse. Such a significant change in the
fidelity always corresponds to a peak on the curve of the FS. Therefore, the
position of the peak of the FS gives us the information of the maximal
effective propagation distance before the output pulse suffers \textit{a
qualitative change} and becomes a completely distorted pulse.

From the information of Fig. 1 and Fig. 2, one can find that there is a
maximal effective propagation distance for the propagation of a smooth
Gaussian pulse in the double-gain abnormal dispersive medium. Within this
maximal effective propagation distance, the shape of the output pulse could
be seen as almost the same as that of the input one, otherwise the output
pulse is completely distorted and has no longer any physical meaning for
transferring the information of the original pulse. Here we would like to
point out that the maximal propagation distance for a smooth Gaussian pulse
depends on not only the dispersive strength of the medium but also the pulse
spectral effective width.

In Fig. 3, we plot the changes of the fidelity and the FS for a smooth
Gaussian pulse propagating through a double-absorptive normal dispersive
medium with $M<0$ in Eq. (\ref{Medium1}). Similar to the cases in Figs. 1(a)
and 1(c), one can find that the fidelity here is still a monotonical,
decreasing function. However, differ from the cases in Fig. 1, there is no
sudden change for the fidelity in Fig. 3(a), so that there is no peak on the
curve of the FS in Fig. 3(b). Note that the value of the FS actually changes
very small in this example [note the very small scale in Fig. 3(b)]. It
means that there is no dramatic change (no qualitative change) for the
output pulse when it propagates inside a normal dispersive medium. The inset
figures in Fig. 3 show that the shape of the pulse is almost the same as the
input one except for the pulse broaden effect. It tells us that the smooth
pulse could always propagate effectively through a normal dispersive medium
without losing its original shape.

Here we would like to point out that, from the above discussions, we can
find that the group velocity is a good concept for the description of a
smooth pulse's propagation inside the normal dispersive medium, while for
the description of a smooth pulse's propagation inside the abnormal
dispersive medium, the group velocity is only useful and effective within a
certain propagation distance before the pulse suffers a significant change.

\subsection{A truncated Gaussian pulse}

Now let us turn to consider the propagation of a \textit{truncated} Gaussian
pulse passing through a dispersive media. Because for a real light pulse,
people believe that there is always a start point and an end point. In our
calculations, the initial wavepacket of a truncated Gaussian pulse in the
time domain could be assumed to be 
\begin{equation}
\Psi \left( 0,t_{1}\right) =\left\{ 
\begin{array}{c}
A_{0}\exp \left[ -\frac{t_{1}^{2}}{2T_{0}^{2}}\right] \exp [-i\omega
_{0}t_{1}],\text{\ }|t_{1}|\leq \xi \\ 
0,\text{ \ \ \ otherwise}\ \ \ \ \ \ \ \ \ \ \ \ \ \ \ \ \ \ \ 
\end{array}%
\right.  \label{Input2}
\end{equation}%
at the incident end $z=0$, where $\xi $ is a truncation parameter of the
Gaussian pulse, here we always take $\xi >2T_{0}$ for a weak truncation.
Although the pulse's shape is similar to the previous case, there are two
special points, a sudden switch-on point at $t_{1}=-\xi $ and a sudden
switch-off point at $t_{1}=\xi $. In this situation, the corresponding pulse
spectral wavepacket in the frequency domain could be numerically obtained by
a Fourier transformation via Eq. (\ref{inputspect}). Using Eq. (\ref%
{Spectrum1}), we can also readily find the pulse spectral wavepacket at any
position inside dispersive media. Similar to the above discussion of a
smooth Gaussian pulse, we can also obtain the change of the fidelity $F(0,z)$
for a truncated Gaussian pulse. From the above discussions, since we have
already known that a sudden change in the fidelity always corresponds to a
peak in the change of the FS, in the following discussion we only calculate
the change of the FS for a truncated Gaussian pulse inside the dispersive
medium and then we discuss the propagation properties of the output pulse.
In the below, we still take $\omega _{0}/2\pi =3.5\times 10^{14}$Hz, and
pulse temporal width $T_{0}=1.2\mu $s.

In Fig. 4(a), we plot the changes of the FSs for different truncated
Gaussian pulses propagating inside a double-absorptive normal dispersive
medium with $M<0$ in Eq. (\ref{Medium1}). From Fig. 4(a), we find that there
exists a dramatic change in the FS curve for a truncated Gaussian pulse at a
certain propagation distance. From the curves a to d, we find that the
positions of the FS's peaks for different truncated pulses are strongly
dependent on the truncation parameter $\xi $. For the small value of $\xi $
(i. e., the strong truncation on the pulse, see Eq. (\ref{Input2})), it is
found that the critical propagation distance (corresponding to the FS's
peak) is very short, for example, see the curve a in Fig. 4(a). As the
truncation of the pulse becomes weaker and weaker, i. e., the value of $\xi $
increases, it is seen that the peak of the FS appears at a longer
propagation distance. Curve e displays the case for a smooth Gaussian pulse.
By comparing the curves a, b, c, and d with the curve e, we find that for a
smooth pulse inside the normal dispersive medium there is no qualitative
change, while for a truncated pulse there is always a qualitative change and
the position of such a FS's peak does strongly depend on the truncation
parameter of the pulse. We suggest that \textit{such an essential difference
between the smooth and the truncated pulses may be tested experimentally.}

In order to understand such a dramatic change in the FS for a truncated
Gaussian pluse inside the normal dispersive medium, we plot Fig. 4(b) to
show the temporal normalized profiles at different propagation distances.
Each figure corresponds to the position denoted on the curve b in Fig. 4(a).
From Fig. 4(b) and the curve b in Fig. 4(a), we find that when the FS
initially changes small, the pulse profile is almost preserved although it
propagates through the normal dispersive medium with a slow group velocity.
However, we notice that there is a sudden switch-on at the time $\tau =-\xi $
and a sudden switch-off at $\tau =\xi $ on the input pulse in the moving
time frame; as the propagation distance increases, such small abrupt changes
in the temporal pulse gradually dominates in the normalized profiles. From $%
z=0$ to $z=35$cm in Fig. 4(b), we see that the Gaussian-shaped peak actually
propagates at the speed of the slow group velocity, while the sudden changes
always occur at the moving times $\tau =-\xi $ and $\tau =\xi $ (i. e., at
the speed of light in vacuum, see our definition of the moving time frame in
section II). Near the position of the FS's peak, the pulse is mainly
dominated by the distortion due to the sudden switch-on and -off. After the
critical propagation distance, the pulse is completely dominated by the
distortion (see the figure for the case of $z=50$cm). Therefore the
qualitative change for a truncated Gaussian pulse inside a normal dispersive
medium comes from the strong distortion near the abrupt changes of the pulse
profile.

Next we consider that such a truncated Gaussian pulse propagates through the
above double-gain abnormal dispersive medium. Figure 5(a) shows the change
of the FS under different truncation parameters. Similar to the cases in
Fig. 4(a), the change of the FS inside the abnormal dispersive medium
strongly depends on the value of the truncation parameter. For the strongly
truncated Gaussian pulse, the position of the FS's peak occurs at a short
propagation distance. For example, see the curve a for the pulse with $\xi
=4T_{0}$ in Fig. 5(a), the FS's peak is nearly located at $z=7.5$cm; while
for the curve b for the pulse with $\xi =5T_{0}$, the peak is located near $%
z=10.5$cm; the curve d is for the case with $\xi =12T_{0}$, which
corresponds to the case of a very weak truncated Gaussian pulse. For a
comparison, we also plot the curve e for a smooth Gaussian pulse in such an
abnormal dispersive medium. Clearly, the position of the FS's peak of the
truncated Gaussian pulse is always smaller than that of a smooth Gaussian
pulse inside the abnormal dispersive medium. This indicates that the
effective propagation distance for a truncated Gaussian pulse is shorter
than that for a smooth pulse; and it becomes shorter and shorter for a
stronger truncation (i.e., a smaller $\xi $).

Figure 5(b) plots the normalized pulse profiles at different positions on
the curve b in Fig. 5(a). For a comparison, we use the dashed curve to
denote the initial truncated Gaussian pulse, see Fig. 5(b). We find that the
peak of the truncated pulse is advanced with the superluminal group velocity
and the temporal shape is nearly preserved at these positions z=5cm and
z=7.5cm. Unlike the cases in Fig. 2, due to the effect of the sudden
switch-on and -off of the truncated pulse, the pulse is gradually distorted
at the beginning propagation region (within the effective propagation
distance); then after the FS' peak, the pulse is completely distorted. This
is why the truncated Gaussian pulse has a smaller effective propagation
distance.

From the above discussion, in terms of the change of the FS for the
truncated pulse, we therefore have that, inside the normal dispersive medium
the group velocity is only valid before the FS's peak of a truncated
Gaussian pulse happens, and inside the abnormal dispersive medium the
effective propagation distance for the validity of the group velocity is
much smaller than that of a smooth pulse.

\section{Summary}

In this work, we have used the concepts of the fidelity and the FS to
describe the pulse propagation inside the dispersive media. We have found
that both the fidelity and the FS are very useful for determining whether
the pulse is completely distorted or not and confirming whether the group
velocity of the pulse is physical meaningful or not. It has been found that
there is a dramatic change in the changes of both the fidelity and the FS
for a smooth Gaussian pulse propagating inside an abnormal dispersive
medium, and the maximal effective propagation distance before the FS's peak
is dependent on the dispersion strength of the medium, and after the FS's
peak the pulse is completely distorted; however when a smooth Gaussian pulse
propagates inside a normal dispersive media, there is no dramatic change in
the changes of the fidelity and its FS and therefore the pulse could always
keep its original shape except for the pulse broaden effect. For a truncated
Gaussian pulse, there always exists a qualitative change on the FS inside
both the abnormal and normal dispersive media, and the maximal effective
propagation distance in this case is strongly dependent on both the pulse
itself (such as the truncation parameter) and the dispersive strength of the
medium. We have found that, under the same medium's parameters, in terms of
the position of the FS's peak, the effective propagation distance for a
truncated pulse inside the dispersive medium is always much smaller than
that of a smooth pulse. These results provide us a possible chance to
analyze the pulse distortion and to determine the effective region of the
pulse group-velocity description, and thus give us a deep understanding on
the pulse propagation inside the dispersive media.

\begin{acknowledgments}
This work was supported by NSFC (No. 10604407) and CUHK 2060360.
\end{acknowledgments}

\newpage

\begin{center}
{\LARGE FIGURE CAPTIONS}
\end{center}

FIG. 1. Changes of the fidelity (a, c) and the FS (b, d) of a smooth
Gaussian pulse inside an abnormal dispersive medium under different medium's
parameters. In (a) and (b), different curves correspond to different $M$;
and in (c) and (d), different curves correspond to different $\Delta $.
Other parameters are $M/2\pi =22.62$Hz, $\Delta /2\pi =1.35$MHz, $\gamma
/2\pi =0.46$MHz, and the carried frequency $\omega _{0}/2\pi =3.5\times
10^{8}$MHz.

FIG. 2. Changes of the normalized intensities of smooth Gaussian pulses
propagating through an abnormal dispersive medium at different positions on
the curve A in Fig. 2(d).

FIG. 3. Changes of the fidelity (a) and the FS (b) of a smooth Gaussian
pulse inside a normal dispersive medium. Inset figures show the normalized
pulse profiles at different positions. Other parameters are the same as in
Fig. 2 except for $M/2\pi =-22.62$Hz.

FIG. 4. (a) Change of the FS for different truncated Gaussian pulses inside
a normal dispersive medium with different truncated parameters: curve a for $%
\xi =4T_{0}$, curve b for $\xi =5T_{0}$, curve c for $\xi =6T_{0}$, curve d
for $\xi =7T_{0}$, and cure e for the smooth Gaussian pulse. Figs. (1) to
(6) show the nomalized intensities of the truncated Gaussian pulse at
different positions on curve b in Fig. (a). The arrows in Figs. (2) and (3)
indicate the sudden changes. The medium's parameters are the same as in Fig.
4.

FIG. 5. (a) Change of the FS for different truncated Gaussian pulses inside
an abnormal dispersive medium with different truncated parameters: curve a
for $\xi =4T_{0}$, curve b for $\xi =5T_{0}$, curve c for $\xi =7T_{0}$,
curve d for $\xi =12T_{0}$, and curve e for the untruncated Gaussian pulse.
Figs. (1) to (4) show the nomalized intensities of the truncated Gaussian
pulse at different positions on curve b in Fig. (a). Te dashed curves in
Figs. (1)-(4) denote the initial truncated Gaussian pulse.

\newpage

\FRAME{ftbpFU}{3.4065in}{5.0341in}{0pt}{\Qcb{FIG.1}}{\Qlb{fig:FIG01}}{%
fig1.eps}{\special{language "Scientific Word";type
"GRAPHIC";maintain-aspect-ratio TRUE;display "USEDEF";valid_file "F";width
3.4065in;height 5.0341in;depth 0pt;original-width 2.4543in;original-height
3.6391in;cropleft "0";croptop "1";cropright "1";cropbottom "0";filename
'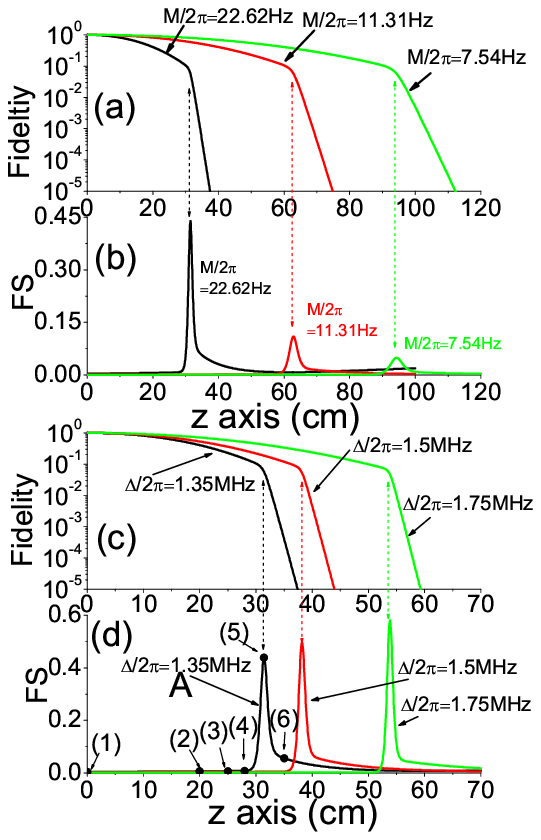';file-properties "XNPEU";}}

\newpage

\FRAME{ftbpFU}{3.6824in}{2.7112in}{0pt}{\Qcb{FIG.2}}{\Qlb{fig:FIG02}}{%
fig2.eps}{\special{language "Scientific Word";type
"GRAPHIC";maintain-aspect-ratio TRUE;display "USEDEF";valid_file "F";width
3.6824in;height 2.7112in;depth 0pt;original-width 2.7198in;original-height
1.9951in;cropleft "0";croptop "1";cropright "1";cropbottom "0";filename
'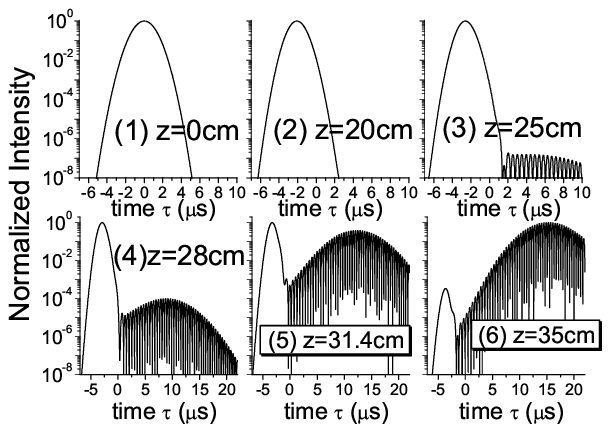';file-properties "XNPEU";}}

\newpage

\FRAME{ftbpFU}{3.4454in}{3.8692in}{0pt}{\Qcb{FIG.3}}{\Qlb{fig:FIG03}}{%
fig3.eps}{\special{language "Scientific Word";type
"GRAPHIC";maintain-aspect-ratio TRUE;display "USEDEF";valid_file "F";width
3.4454in;height 3.8692in;depth 0pt;original-width 2.8297in;original-height
3.1808in;cropleft "0";croptop "1";cropright "1";cropbottom "0";filename
'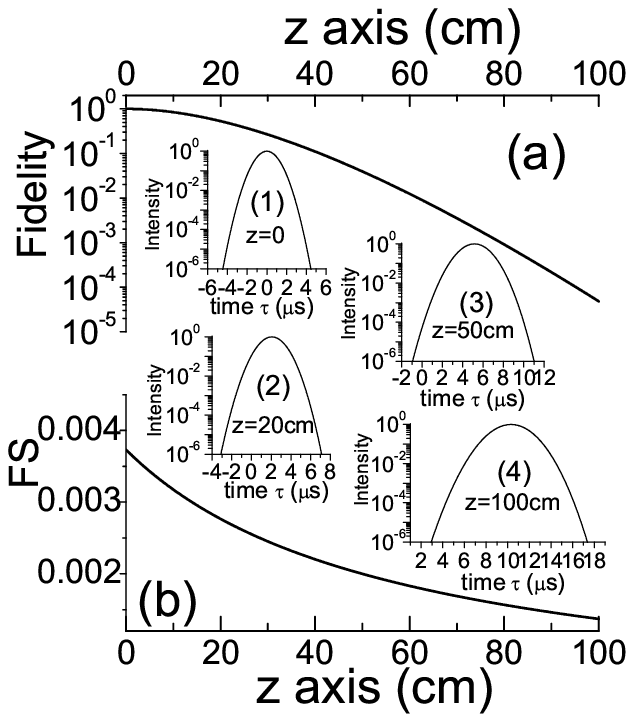';file-properties "XNPEU";}}

\newpage

\FRAME{ftbpFU}{3.8475in}{4.7167in}{0pt}{\Qcb{FIG.4}}{\Qlb{fig:FIG04}}{%
fig4.eps}{\special{language "Scientific Word";type
"GRAPHIC";maintain-aspect-ratio TRUE;display "USEDEF";valid_file "F";width
3.8475in;height 4.7167in;depth 0pt;original-width 3.2638in;original-height
4.0075in;cropleft "0";croptop "1";cropright "1";cropbottom "0";filename
'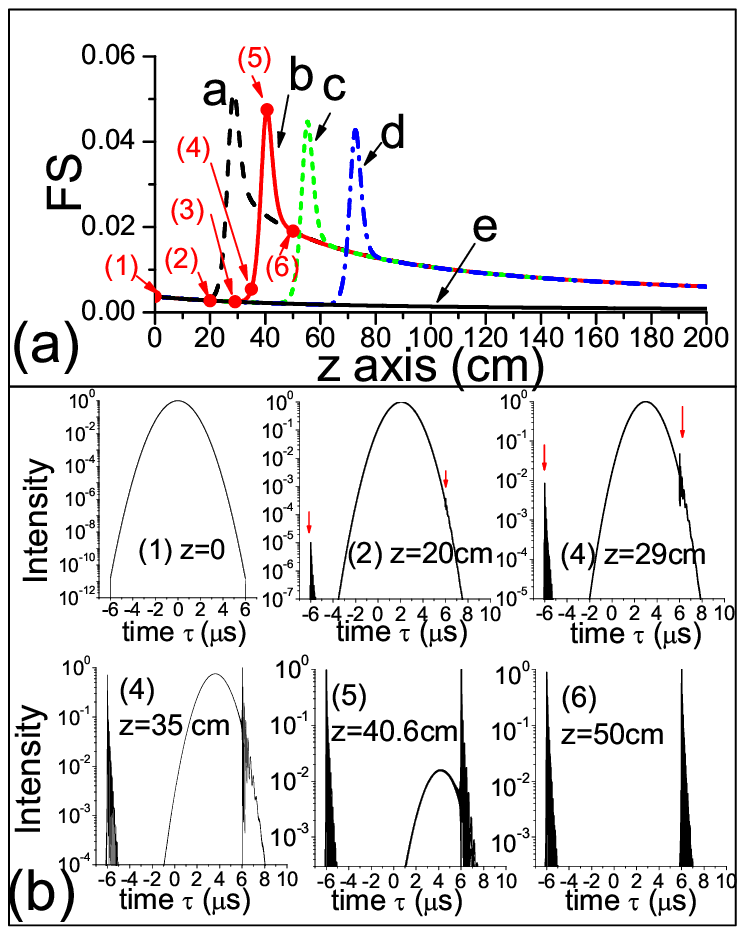';file-properties "XNPEU";}}

\newpage

\FRAME{ftbpFU}{3.7317in}{4.3483in}{0pt}{\Qcb{FIG.5}}{\Qlb{fig:FIG05}}{%
fig5.eps}{\special{language "Scientific Word";type "GRAPHIC";display
"USEDEF";valid_file "F";width 3.7317in;height 4.3483in;depth
0pt;original-width 2.8297in;original-height 3.2085in;cropleft "0";croptop
"1";cropright "1";cropbottom "0";filename '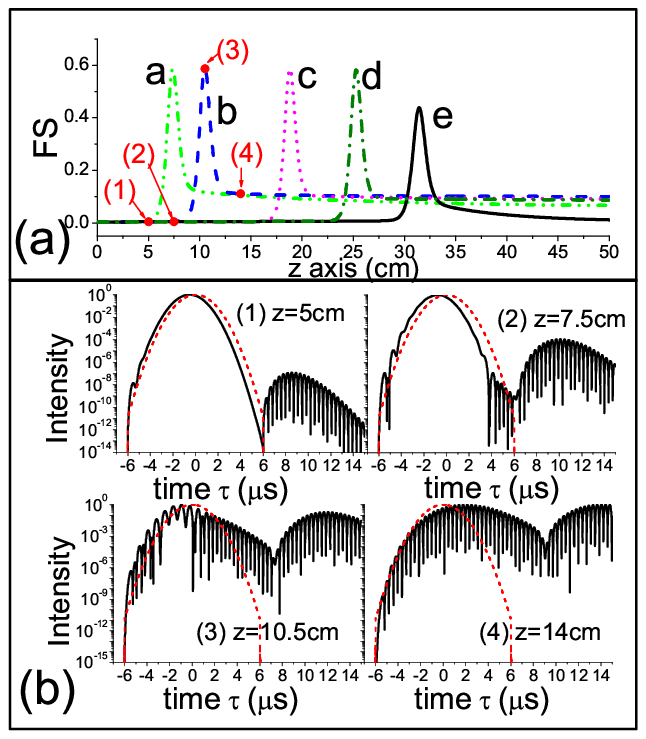';file-properties
"XNPEU";}}

\end{document}